# Sub-optimum Signal Linear Detector Using Wavelets and Support Vector Machines

Jaime Gomez[1]
Ignacio Melgar[2]
Juan Seijas[3]
Diego Andina[4]

Sener Ingeniería y Sistemas, S.A. [1 2 3]
Departamento de Señales, Sistemas y Radiocomunicaciones, Universidad Politécnica de Madrid. [2 3 4]
Escuela Politécnica Superior, Universidad Autónoma de Madrid [1]
jaime.gomez@ii.uam.es
ignacio.melgar@sener.es
seijas@gc.ssr.upm.es

*Abstract:* - The problem of known signal detection in Additive White Gaussian Noise is considered. In previous work, a new detection scheme was introduced by the authors, and it was demonstrated that optimum performance cannot be reached in a real implementation. In this paper we analyse Support Vector Machines (SVM) as an alternative, evaluating the results in terms of Probability of detection curves for a fixed Probability of false alarm.

## 1 Introduction

Wavelet transform is one of the most successful methods in signal de-noising processes.
Let **s** be a sampled pulse with energy contained in a certain range of frequencies, and let **n** be sampled noise alone. The wavelet coefficients for both, pulse plus noise or only noise signals, can be computed using Subband Coding Algorithm [4]. The wavelet coefficients with greater signal to noise ratio (SNR) will be those which correspond to frequency scales where signal energy is concentrated.

In [10] a signal detection algorithm is proposed. It is based on comparing the component with maximum absolute value of its Discrete Wavelet Transform (DWT) coefficients, for a given scale, with a certain threshold. The Probability of false alarm ($P_{fa}$) is set to a defined value. The threshold ($V_T$) which fits with Pfa requirements is obtained using Montecarlo estimation and a set of noise DWT coefficients. Threshold depends on the wavelet scale used for comparison. The same process is used to obtain Probability of detection ($P_D$), now with a set of pulse DWT coefficients. Only one component of the DWT vector is used, therefore a big amount of information is being rejected.

In [6] a new algorithm called Linear-Detector is introduced. It compares the scalar product of two vectors with the threshold, instead of the maximum component of the DWT domain. One of the two vectors is the coefficients of the DWT applied to the signal for a given scale. The other one (vector **a**) is obtained as the one which achieves optimum detection performance in terms of Probability of detection for a given Probability of false alarm. The **a**–finding algorithms show that its optimum value depends on the SNR, which is unknown in input signal. Therefore, optimum-performance linear detector real implementation is not possible. Theoretical maximum Probability of detection curves are also presented in [6], so it is possible to evaluate any method to reach a non-SNR dependent **a** vector, obtaining its $P_D$ versus SNR curve and comparing it with the theoretical optimum.

In this paper we study linear Support Vector Machines as an alternative algorithm to overcome input SNR dependence. Linear SVM architecture computes scalar product and threshold comparison internally, so the dual-state output is considered as the final binary-detection result. The **a**-obtaining algorithm now corresponds to the training phase of the SVM. A fine adjustment of the training parameters is necessary, because a predefined $P_{fa}$ value has to be reached. Figure 1 shows the data-flow schematic.

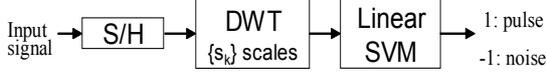

Fig.1: Detection scheme under study

## 2 Wavelet transform

Almost all existing signals can be expressed by a wavelet transform. Wavelets are generated by the scale and translation of a single prototype function called wavelet mother:

$$\psi_{s,\tau}(t) = \frac{1}{\sqrt{|s|}} \psi\left(\frac{t-\tau}{s}\right), \quad (1)$$

where $\psi$ is the mother wavelet, s is the scaling factor and $\tau$ is the translation factor. They are building blocks of wavelet transform for different scales and translations, just as trigonometric functions of different frequencies are building blocks of Fourier transform [8].

In the case of discrete wavelet transform (DWT), $\tau$ and s also take discrete values, given by

$$s = s_0^m, \quad (2)$$

$$\tau = n\tau_0 s_0^m, \quad (3)$$

$$\psi_{m,n}(t) = s_0^{-m/2} \psi(s_0^{-m} t - n\tau_0), \quad (4)$$

A particular class of wavelets are orthonormal wavelets which are linearly independent, complete and orthogonal. This means that there is no "redundant" data from the original signal in more than one wavelet. In [3] Daubechies developed conditions under which wavelets form orthonormal bases. Thus the Discrete Wavelet Coefficients are the inner products of the signal and wavelet function. That is:

$$f(t) = \sum_{m,n} \psi_{m,n}(t) <\psi_{m,n}(t), f(t)>, \quad (5)$$

Multiresolution analysis allows us to look a signal at different scale by "zooming in " or "zooming out"; that is, an approximation of a given signal at low resolution can go to an approximation at immediate higher resolution just by adding some "details" information.

In [4] Mallat developed a fast wavelet algorithm based on the pyramid algorithm of Burt and Adelson [2]. The basic components in each stage of the pyramid are two analysis filters: a low-pass filter **h** and high-pass filter **g**, and a decimation by two operation. As it can be easily observed, because of the Heisenberg uncertainty principle; DWT offers high time resolution for low scales (high frequencies) and high frequency resolution for high scales (low frequencies).

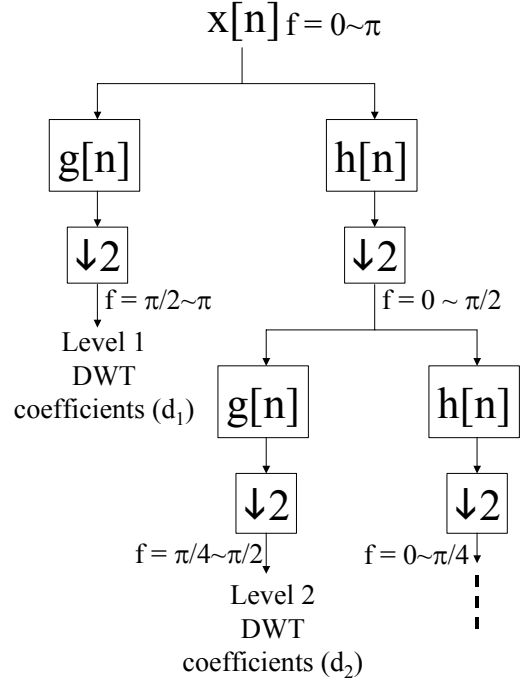

Fig.2 The dyadic decomposition algorithm

## 3 Linear-Detector

### 3.1 Algorithm description

In [6] the linear-detector algorithm is described. Let us consider **s** as a N-length vector containing a sampled signal. The signal may consist on a known pulse inside noise with a certain SNR or just noise. As described in the introduction, Discrete Wavelet Transform is computed for a certain i-th scale, being $\mathbf{d_i}$ the M-length wavelet coefficients vector:

$$\mathbf{d}_i = DWT_i(\mathbf{s}) \quad (6)$$

A new M-length vector, denoted by **a**, is introduced. Scalar product between **a** and $\mathbf{d_i}$ is computed and compared with the threshold for pulse detection.

$$\upsilon = \sum_{k=1}^{M} \mathbf{d}_i[k] \cdot \mathbf{a}[k] \quad (7)$$

The algorithm described in next section shows how the value of **a** which ensures optimum detecting performance is SNR dependent. With the introduction of a SVM, a non SNR-dependent sub-optimum **a** vector and a threshold which satisfies $P_{fa}$ requirements are obtained as the result of the training phase.

### 3.2 Optimum-finding algorithm

In [6] the algorithm used to determine the presence of a known signal inside additive white Gaussian noise (AWGN) is described. Nevertheless, we include the complete development for paper completeness.

Using vector notation for signal representation, let **v** be defined as a $2^N$ length vector containing a sampled pulse in the presence of AWGN. Noise distribution is zero-mean and $\sigma_n^2$ variance. For the true hypothesis ($H_1$) vector **v** is the addition of sampled noise and sampled signal with a certain signal-to-noise ratio (SNR). For the false hypothesis ($H_0$) signal is not present, thus **v** contains only noise. Therefore **v** can be expressed, using natural scale, as follows:

$$\mathbf{v}|_{H_0} = \mathbf{n}, \quad (8)$$

$$\mathbf{v}|_{H_1} = 10^{\frac{SNR}{20}} \sigma_n^2 \hat{\mathbf{s}} + \mathbf{n}, \quad (9)$$

where $\hat{\mathbf{s}}$ is the sampled pulse sequence with power normalized to 1, **n** is the sampled AWGN and SNR is expressed in dB units.

Using Subband Coding algorithm [4] we obtain $\mathbf{d}_{i(s)}$ vector as the i-th level detail wavelet coefficients of $\hat{\mathbf{s}}$. $\mathbf{d}_{i(s)}$ vector length M is given by

$$M = 2^{N-i}. \quad (10)$$

Assuming that high-pass and low-pass filter impulse responses h[n] and g[n] are normalized to unitary energy, that is

$$\sum_{n=0}^{L-1}(h[n])^2 = \sum_{n=0}^{L-1}(g[n])^2 = 1, \quad (11)$$

being L the filter length, then i-th wavelet coefficients of vector **v** can be written as

$$\mathbf{d}_i|_{H_0} = \mathbf{d}_{i(n)}, \quad (12)$$

$$\mathbf{d}_i|_{H_1} = 10^{\frac{SNR}{20}} \sigma_n^2 \mathbf{d}_{i(s)} + \mathbf{d}_{i(n)}. \quad (13)$$

$\mathbf{d}_{i(n)}$ vector corresponds to i-th detail wavelet coefficients of **n** and length given by (10). Each k-th component of $\mathbf{d}_{i(n)}$ is a zero-mean and $\sigma_n^2$ variance gaussian random variable for $L \leq k \leq M$ (when the filter reaches the steady stage).

Let us introduce the variable $\upsilon$. Pulse presence will be asserted if the value of $\upsilon$ is greater than a certain threshold ($V_T$). Now let define $P_{fa}$ and $P_D$ as

$$P_{fa} = \Pr\{\upsilon > V_T / H_0\}, \quad (14)$$
$$P_D = \Pr\{\upsilon > V_T / H_1\}. \quad (15)$$

For a given vector $\mathbf{a} \in \Re^M$, $\upsilon$ is created by

$$\upsilon = \sum_{k=L}^{M} \mathbf{d}_i[k]\mathbf{a}[k]. \quad (16)$$

So, $\upsilon$ is a gaussian random variable with $N(\eta_\upsilon, \sigma_\upsilon)$ normal distribution, where

$$\eta_\upsilon|_{H_0} = E\{\upsilon\}|_{H_0} = \sum_{k=L}^{M}\mathbf{a}[k]E\{\mathbf{d}_{i(n)}[k]\} = 0, \quad (17)$$

$$\eta_\upsilon|_{H_1} = E\{\upsilon\}|_{H_1} = \sum_{k=L}^{M}\mathbf{a}[k]E\left\{10^{\frac{SNR}{20}}\sigma_n^2\mathbf{d}_{i(s)}[k]\right.$$
$$\left. + \mathbf{d}_{i(n)}[k]\right\} = 10^{\frac{SNR}{20}}\sigma_n^2 \sum_{k=L}^{M}\mathbf{a}[k]E\{\mathbf{d}_{i(s)}[k]\}, \quad (18)$$

and

$$\sigma_\upsilon^2|_{H_0} = \sigma_\upsilon^2|_{H_1} = \sigma_n^2 \sum_{k=L}^{M}(\mathbf{a}[k])^2 +$$
$$2\sum_{k=L}^{M-1}\left[\sum_{j=k+1}^{M}E\{\mathbf{a}[k]\mathbf{a}[j]\mathbf{d}_{i(n)}[k]\mathbf{d}_{i(n)}[j]\}\right]. \quad (19)$$

The second term in the sum of (19) is negligible with respect to the first term. Therefore we simplify that expression to the following one:

$$\sigma_\upsilon^2|_{H_0} = \sigma_\upsilon^2|_{H_1} = \sigma_n^2 \sum_{k=L}^{M}(\mathbf{a}[k])^2. \quad (20)$$

As it is shown in (17), (18) and (20), for a given $P_{fa}$ and vector **a** with at least one non-zero component, the associated $V_T$ can be obtained. If SNR is also considered then $P_D$ can be computed as well. Therefore it is possible to reach maximum detection performance of the scheme proposed in (16) only by maximizing $P_D$ as a function of SNR and **a**, for the previously fixed $P_{fa}$.

The proposed algorithm can be extended to improve probability of detection. Let us suppose detail wavelet coefficients of a sampled pulse for a set of K scales $B = \{i_k \mid 1 \leq i_k \leq N\}$ are computed. Then all of them are concatenated on a single vector $\mathbf{d}_B$, created as follows:

$$\mathbf{d}_B = [\mathbf{d}_{i_1}, \mathbf{d}_{i_2}, \ldots \mathbf{d}_{i_K}]. \quad (21)$$

Now let us use the detection process proposed in (16) with $\mathbf{d}_B$. For a given $\mathbf{a}$ vector with same length as $\mathbf{d}_B$, (18) and (20) show $\sigma_\upsilon^2$ (and consequently $P_{fa}$) do not depend on set B, but $\eta_\upsilon$ and $P_D$ do. Therefore, if we only include on set B the wavelet coefficients for the scales where the sampled pulse has high amplitude in any of its components, then $\eta_\upsilon$ and $P_D$ will be increased. That means the chosen $i_k$–th scale wavelet coefficients will correspond to the discrete frequency ranges $[\pi/2^{i_k}, \pi/2^{i_k-1}]$ where the pulse has greater energy.

### 3.3 Support Vector Machines

Support Vector Machine (SVM) is a classification tool introduced by V. Vapnik in 1995 [11]. Since then, it has been used in a variety of problems with excellent results.

A SVM separates two classes using an optimum discriminator hyperplane so to maximize a convex quadratic objective function.

$$L_D = \sum_{i=1}^{l} \alpha_i - \frac{1}{2} \sum_{i=1}^{l} \sum_{j=1}^{l} \alpha_i y_i \alpha_j y_j (x_i \bullet x_j), \quad (22)$$

where x are the training patterns, y are their class and $\alpha$ are the patterns weight coefficients. These coefficients are found during training.

This simple algorithm has remarkable properties: there is one solution only (no local minima); SVM parameters are few and easy to handle; data separation is performed in a very high dimensional feature space, making it much easier; new features are not calculated explicitly, so there is no complexity increase regardless of the use of high dimensional spaces; expected noise figures are introduced in the training algorithm, upgrading robustness; generalization capability is outstanding, despite the high dimensional space.

To generate a valid comparison between the theoretical limits described in the previous section and the numerical results, only the linear kernel can be applied. Note that in the linear case the separating hyperplane can be calculated explicitly and, therefore, both theoretical and numerical representations have a similar analytic formula.

$$w = \sum_{i=1}^{l} \alpha_i y_i x_i, \quad (23)$$

$$f(x) = (w \bullet x) + b. \quad (24)$$

After the training, the optimum values for b and w are calculated. Then, b corresponds to the algorithm threshold, and vector w corresponds to vector $\mathbf{a}$, as they were defined on section 3.1.

The training process needs two opposite-class sets. The first one (positive class) contains the wavelet transform representation of pulse signals combined with noise for different SNR ratios form 0 to –15 dB, identically distributed. We chose that SNR interval because it is the most significant in our experiments. The second one (negative class) contains wavelet transform representation for noise alone. Note that even in the case both sets had the same size, the training process does not require equilibrium, i.e., false negatives (probability of detection) and false positives (probability of false alarm) do not have the same constraints. In our experiments we required that probability of false alarm be $10^{-3}$, so we had to force false positives errors to have more weight on the objective function. For that purpose we used different combinations of soft-margin parameter C (we used C+ and C– version) to achieve the desired $P_{fa}$ on a noise set. Afterwards, the probability of detection is calculated for different signal-plus-noise subsets with a specific SNR.

### 3.4 Algorithm complexity

The computational load for both theoretical and numerical algorithms depend on the phase.

For the optimum finding coefficients calculation, resources in the theoretical algorithm are spent in single SNR optimization function. Using the SVM, only one execution is needed for all SNR values. The SVM training algorithm has complexity $O(N^2)$ where N is the number of important patterns. Using a linear kernel the number of important patterns in training is reduced drastically, decreasing needed computational resources. This process is done only once, during the engineer system development and its performance is not critical.

For the operative phase, both theoretical and numerical algorithms have the same analytical expression, and so, they have identical complexity. First, the wavelet transform is calculated, and then the linear function is applied. Being N the initial vector size, and M the wavelet filter size, the number of multiply-add operations performed in the wavelet transform is 2MN. The number of multiply-add operations needed in the linear function calculation is N. Therefore, the computational resources needed for both complete algorithms are basically the same as for the wavelet transformation calculation, so the complexity remains O(MN).

## 4 Experimental results

The data used to check sub-optimal performance had the following features: chirp pulse, 1024 samples (see Fig.3); mother wavelet Daubechies 5, using $d_3$, $d_4$, $d_5$ and $d_6$ wavelet coefficients. For all experiments we used white Gaussian noise with zero mean and deviation equals one, and $P_{fa} = 10^{-3}$.

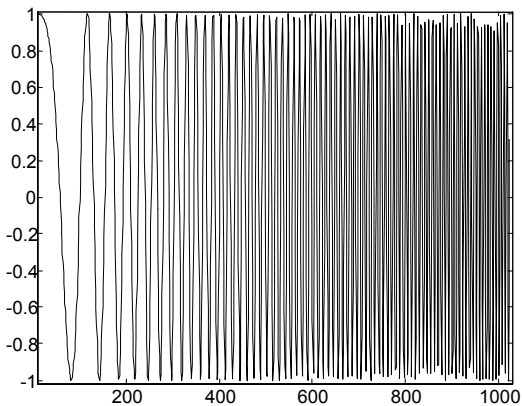

Fig.3: Chirp pulse with 1024 samples.

It can be observed in Fig.4 that SVM results are very close to theoretical limits in basic wavelet scales. Nevertheless, as the input vector increases, differences grow. This effect was expected, because when the quantity of information to be integrated increases, SNR scenario representations have greater differences.

In the sup-optimum algorithm we can still use together more than one scale, more than one wavelet transform, or even more than one kind of filter. The more non-redundant information you use, the better probability of detection you will get. There is no theoretical limit to the quantity of information you can draw together.

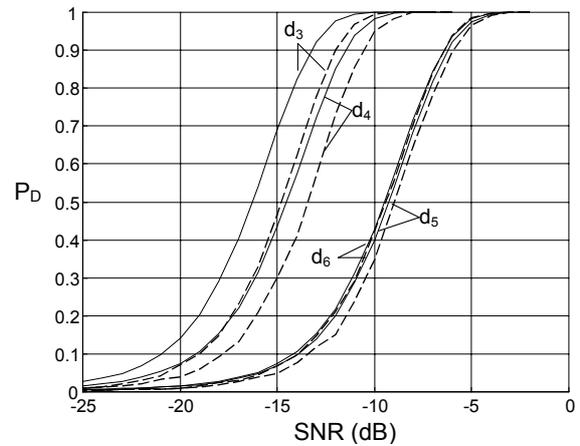

Fig.4: Comparison of different wavelet scales using theoretical limits (solid lines) and SVM approach (dashed lines). $P_{fa}$ is always set to $10^{-3}$.

In Fig.5 the differences in theoretical and SVM results for a simple multiple-source representation are shown. Using the same basic parameters as in the previous experiments we concatenated the results of applying $d_3$, $d_4$, $d_5$ and $d_6$ wavelet coefficients in one single vector. While the probability of detection is increased compared to the best single scale, SVM results are a bit lower, but they are still far better than single scale results.

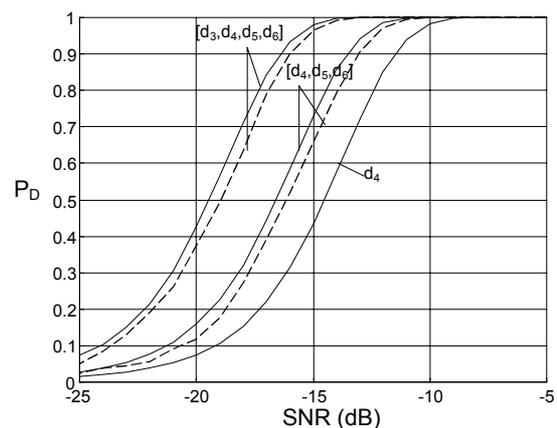

Fig.5: Comparison of two multi-scale vector, (leftmost is concatenation of $d_3$, $d_4$, $d_5$, and $d_6$, and the following is concatenation of $d_4$, $d_5$, and $d_6$), for theoretical limits (solid lines) and SVM approach (dashed lines). Also, a single scale $d_4$ theoretical limit is shown for easier comparison. $P_{fa}$ is always set to $10^{-3}$.

## 5 Conclusions

The algorithm introduced in [6] gave theoretical reachable optimum performance, but it could not be implemented in a real system because its applicability depends on a specific SNR value, which is unknown in the signal processor. This paper has shown an alternate way to calculate the optimum Linear-Detector coefficients without the constraints about SNR. For that purpose we used the best machine learning algorithm currently, Support Vector Machines, turning the algorithm practical.

The results show two interesting properties. First, the SVM performance is quite close to the theoretical limits, confirming both that the SVM choice was correct, and that theoretical results are reachable but cannot be beaten. Second, multiple source information continues to be applicable, leaving the possibility of more research on this issue. Note that, even though we used a specific wavelet transform and chirp pulse, there is no constraint in the SVM application about these parameters.

In most generic problems, the use of non-linear kernel usually gives better performance. However, during the experiments it was seen that the best classifier was the one having a linear decision function in input space. Even though linear kernels can be seen as a particular case of non-linear kernels, the fewer parameters in the former formulation give an easier way to find the best solution. Non-linearity does not give better features to solve this problem, and therefore it is not useful.

## 6 Acknowledgements
This project is funded by Sener Ingeniería y Sistemas, in the frame of the Aerospace Division R&D program.